\documentclass[12pt]{article}
\begin{document}
\newcommand{\beq}{\begin{equation}}
\newcommand{\eeq}{\end{equation}}
\newcommand{\beqn}{\begin{eqnarray}}
\newcommand{\eeqn}{\end{eqnarray}}
\newcommand{\dpf}{\displaystyle\frac}
\newcommand{\no}{\nonumber}
\newcommand{\ep}{\epsilon}
\begin{center}
{\Large Relating Friedmann equation to Cardy formula in universes with cosmological
constant}
\end{center}
\vspace{1ex}
\centerline{\large Bin
Wang$^{a,b,}$\footnote[1]{e-mail:binwang@fma.if.usp.br},
\ Elcio Abdalla$^{a,}$\footnote[2]{e-mail:eabdalla@fma.if.usp.br}
\ and Ru-Keng Su$^{c,}$\footnote[3]{e-mail:rksu@fudan.ac.cn}}
\begin{center}
{$^{a}$ Instituto De Fisica, Universidade De Sao Paulo, C.P.66.318, CEP
05315-970, \\ Sao Paulo, Brazil \\
$^{b}$ Department of Physics, Shanghai Teachers' University, P. R. China \\
$^{c}$ CCAST (World Lab), P.O.Box 8730, Beijing, P. R. China \\ and Department of
Physics,
Fudan University, Shanghai 200433, P. R. China}
\end{center}
\vspace{6ex}
\begin{abstract}
A relation between the Friedmann equation and the Cardy formula has been found for de
Sitter closed and Anti de Sitter flat universes. For the remaining (Anti) de Sitter
universes the arguments fail, and we speculate whether the general philosophy of
holography can be satisfied in such contents.
\end{abstract} \vspace{6ex}
\hspace*{0mm} PACS number(s): 04.70.Dy, 98.80.Cq
\vfill
\newpage
Twenty years ago, Bekenstein argued that for any isolated physical system of energy $E$
and size $R$, the usual thermodynamic entropy is bounded by $S \leq S_B =2\pi RE$ [1]. By
choosing $R$ to be the particle horizon, he gave a prescription for a cosmological
extension of the bound [2]. In view of the example of black hole entropy, an influential
holographic principle was put forward recently, which suggests that the maximal entropy
is bounded by the area of the spacelike surface enclosing a certain region of space [3].
For systems of limited gravity, Bekenstein's bound implies the holographic bound. The
extension of the holographic bound to cosmological settings was first addressed by
Fischler
and Susskind (FS)[4]. They have shown that for flat and open
Friedmann-Lemaitre-Robertson-Walker(FLRW) universes the area of the particle horizon
should bound the entropy on the backward-looking light cone. However violation of FS
bound was found for closed FLRW universes. Various different modifications of the FS
version
of the holographic principle have been raised subsequently [5-11]. In addition to the
study of holography in homogeneous cosmologies, attempts to generalize the holographic
entropy bound to a generic realistic inhomogeneous cosmological setting were carried out
in [12,13]. 

Despite different versions of the holographic entropy bounds constructed in cosmology,
the
microscopic understanding of these entropy bounds is still lacking. Light on this problem
was first shedded in [14] by showing that in the FLRW flat universe the horizon area can
be
calculated by state counting. The problem was also discussed in an interesting
recent paper by Verlinde [15]. He revealed that when a universe-size black hole can be
formed, the holographic entropy bound is satisfied and there appears a deep and
fundamental connection between the holographic principle, the entropy formula for the
conformal field theory (CFT) and the Friedmann equations for a radiation dominated closed
universe. This finding is remarkable because it not only relates the entropy bound to the
Cardy formula showing the statistical meaning of the bound, but also relates the
Friedmann
equation describing the bulk to the Cardy formula on the boundary and unifies the
Bekenstein
bound [1] and Hubble bound [7] in a cosmological context. The implicit quantum
corrections to Cardy formula has also been given in [16]. Verlinde's generalization of
Cardy formula was subsequently challenged by weak-coupling, high temperature CFT
calculations [17]. Extending the causal entropy bound (CEB) to arbitrary dimensions and
checking it against CFT calculations, Brustein et al [18] claimed that CEB can pass the
CFT test and evade the criticism in [17]. They also argued that for a large class of
models including high temperature weakly and strongly coupled CFTs with AdS duals, CEB is
equivalently to a purely holographic entropy bound proposed in [15]. A very recent study
showed that even for free CFTs in dimensions $D=4,6$, there are specific cases where
Cardy formula can still hold [19], which supports Verlinde's result.

It is of interest to
generalize Verlinde's discussion to a broader class of universes including a cosmological
constant and investigate whether his result is universally true. 
We will suppose that a de Sitter or an Anti-de Sitter (AdS) universe is occupied by a
universe-size black hole. For the AdS universe occupied by a black hole of the universe
size, according to the AdS/CFT duality conjecture [20], the system is associated to a
strongly coupled CFT residing on the conformal boundary. For the de Sitter spacetime, the
close relation between entropy associated with cosmological de Sitter horizon and CFT has
been revealed [21]. Recent studies proved that de Sitter spacetime can be embedded in the
AdS spacetime and presented the nature of strongly coupled CFT in de Sitter specetime as
well [22]. In this paper we will
try to build up the relation between the Friedmann equation and entropy bounds and
present an
intriguing resemblance of Cardy formula in de Sitter and AdS universes.

A. De Sitter universes.

First we will consider universes with a positive cosmological constant. The Friedmann
equation is given by  
\beq     
H^2=\dpf{8\pi G}{3}\dpf{E}{V}+\dpf{\Lambda}{3}-\dpf{k}{R^2}
\eeq
where $H=\dot{R}/R$ is the Hubble parameter, $E/V$ represents the energy density,
$\Lambda$ the cosmological constant. $k=1, -1$ or $0$ corresponds to closed, open or flat
cosmological models, respectively. We will concentrate our attention on the closed model
in the following, which expands at first until it reaches a maximal radius, and
subsequently recollapses. Gravity with a cosmological constant has been studied since
long time (see [23]). The general situation, concerning stability and general properties
of theory is such that they cannot be excluded (see introduction of [23]). Anywhere
inside the event horizon, de Sitter space is stable.

In [24], we have shown from a Geroch process that the Bekenstein bound does not change
the
form if the background spacetime has a cosmological constant. Thus in the present
context,
namely that of a closed de Sitter universe with radius $R$ and total energy $E$, the
appropriate Bekenstein bound is still
\beq             
S<S_B
\eeq
where the Bekenstein entropy is $S_B=2\pi ER$.

The Bekenstein bound is only appropriate for systems with limited self-gravity 
satisfying $HR<1$. In a strongly self-gravitating universe with $HR>1$, the possible
formation of a black hole has to be considered and the entropy bound has to be
modified. Using the idea proposed in [7,8], the total entropy of the strongly gravitating
system is bounded by the Hubble bound $S_H=n_H S^H$, where $n_H$ is the number of
cosmological horizons within a given comoving volume divided by a volume of a single
horizon $n_H=V/\vert H\vert^{-3}$; $S^H$ is the entropy within a given horizon
$\dpf{\vert H\vert^{-2}}{4G}$. The appropriately normalized Hubble bound for $HR>1$ takes
the form [15]
\beq       
S_H=\dpf{HV}{2G}.
\eeq

The holographic principle is based upon the idea that the maximal entropy of a volume $V$
is given by the largest black hole that fits inside $V$ [3]. 

A black hole in de Sitter space is of the form
\beq     
ds^2=-f(r)dt^2+f^{-1}(r)dr^2+r^2d\Omega^2_2,
\eeq
where $f(r)=1-\dpf{2MG}{r}-\dpf{r^2}{r_0^2}$, and $r_0=\sqrt{3/\Lambda}$.  There are two
positive roots of the cubic
equation $f(r)=0$ corresponding to black hole horizon $r_+$ and cosmological horizon
$r_c$. As the mass parameter $M$ increases, the black hole radius $r_+$ increases and the
cosmological horizon $r_c$ decreases monotonically. They become equal for maximal value
of $M$, $r_+=r_c=\sqrt{1/\Lambda}$, which is the biggest black hole that can be formed in
de Sitter space.

Following directly from the Friedmann equation (1) for closed universe, we find when the
universe goes from the weakly to strongly self-gravitating phase, the energy sufficient
to form a black hole of the size of the universe, $r_+=R=\sqrt{1/\Lambda}$, is
\beq       
E_{BH}=\dpf{5V}{8\pi G R^2}
\eeq
and the holographic Bekenstein-Hawking entropy of the black hole with universe-size reads
\beq       
S_{BH}=\dpf{5V}{4GR}.
\eeq

With Eqs(2,3,6) at hand, we can rewrite the Friedmann equation (1) in closed
universe in the form
\beq              
S_H^2=\dpf{4}{75}(5S_B S_{BH}-2S_{BH}^2).
\eeq
Writing $S_H$ in terms of the energy $E$, the radius $R$ and the Bekenstein-Hawking
energy $E_{BH}$, Eq(7) becomes
\beq      
S_H=\dpf{2\pi}{5\sqrt{6}}\sqrt{8E_{BH}R(5ER-2E_{BH}R)}.
\eeq
After making the identifications $L_0=5RE, \dpf{c}{6}=8RE_{BH}$, Eq(8) can be
changed to
\beq           
S_H=\dpf{2\pi}{5\sqrt{6}}\sqrt{\dpf{c}{6}(L_0-\dpf{c}{24})}.
\eeq
This is exactly the Cardy formula except for the difference in the numerical pre-factor.
Inside the event horizon every term makes sense [23].

Therefore at the turning point $HR=1$, when the universe develops from a weakly to
strongly gravitating system, there is a possibility of forming a largest black hole
within
the de Sitter closed universe and the correspondence between the Friedmann equation and
the Cardy formula holds at this moment. Eq(9) unifies the Bekenstein bound and Hubble
bound in the de Sitter cosmological context. Along the evolution of the closed universe
it would be no problem to satisfy both Bekenstein bound for weak gravity regime and
Hubble bound for strong gravity regime if (9) holds.

The matching of the Friedmann equation to Cardy formula cannot be realized in open and
flat de Sitter universes, because  those universes always expand and
self-gravity
of the system is always weak. Only Bekenstein entropy bound can be satisfied in those
context. This result agrees to that in [17] where they found that the dual theory does
not hold in weakly coupled CFT's with AdS duals.

B. Anti-De Sitter universes.

Now we start to study the universes with negative cosmological constant. Usually,
cosmologists believe that closed universes collapse, whereas open and flat universes
expand forever. But the situation is not quite so simple. If there is a negative
cosmological constant, the universe collapses independent of whether it is closed, open
or flat. The Friedmann equation in AdS universes is
\beq      
H^2=\dpf{8\pi G}{3}\dpf{E}{V}-\dpf{\Lambda}{3}-\dpf{k}{R^2}
\eeq
where $H$ the Hubble constant, $E/V$ denotes the energy density, $-\Lambda$ is the value
of the negative cosmological constant. $k=1, -1, 0$ corresponds to closed, open or flat
universes, respectively.

From [24] we learnt that the Bekenstein bound
still holds as expressed in (2) in AdS universes when the gravitational self-energy is
limited ($HR < 1$). When the gravitational self-energy becomes strong ($HR>1$), the
Bekenstein
entropy bound breaks down. We need to introduce the Hubble entropy bound Eq(3) in the
spirit
of [7,8] in a strongly self-gravitating universe.

In the AdS space, the black hole has the metric (4), where
$f(r)=1-\dpf{2MG}{r}+\dpf{r^2}{r_0^2}$ and $r_0=\sqrt{3/\Lambda}$ is the AdS radius. For
a large AdS black hole ($r_0\ll r_+ < 2MG$ [25]) we find from $f(r)=0$ that the black
hole radius is
\beq       
r_+=(6MG/\Lambda)^{1/3}- 1/(6GM\Lambda^2)^{1/3}.
\eeq
The relation between the cosmological constant $\Lambda$ and black hole horizon $r_+$
reads
\beq      
\Lambda =6GM/r_+^3-3/r_+^2.
\eeq

For the flat AdS universe $(k=0)$, we suppose the universe is filled with the
universe-size
AdS black hole ($R=r_+$). From (10) we have $HR=1$ if we take $E=M, V=4\pi R^3/3$ and
(12). This is exactly the turning point between the weakly and strongly self-gravitating
system. According to [15], a Cardy formula can be deduced here to unify the Bekenstein
bound and Hubble bound at this turning point. However we cannot directly get the
expression of the energy to form a black hole as that obtained in [15], or else in the de
Sitter case above, as obtained from the Friedmann equation (10).

Following the idea introduced in [7] by Veneziano, we can write the scale of causal
connection instead of the Hubble radius
$H^{-1}$, that is,
\beq         
H^{-1}\rightarrow R\int \dpf{dt}{R}=R\int^R_0 \dpf{dR}{R\dot{R}}\equiv d_H.
\eeq
Using (10) and considering the flat AdS universe is occupied by the universe-size AdS
hole ($R=r_+)$, we get
\beq     
d_H=R^{2/3}(2GE)^{-1/2}(1-R/2GE)^{-1/6}.
\eeq
Substituting (14) into the Hubble bound (3) with $H\rightarrow d_H$, we have 
\beq   
S^2_H=\dpf{8\pi^2 ER^3}{9G}-\dpf{4\pi^2 R^4}{27G^2}.
\eeq
Since the holographic Bekenstein-Hawking entropy of the universe-size AdS black hole is
$S_{BH}=\dpf{A}{4G}=\dpf{\pi R^2}{G}$ and the Bekenstein entropy $S_B=2\pi ER$, (15) can
be rewritten as
\beq  
S^2_H=\dpf{4}{9}S_{BH}(S_B-\dpf{1}{3}S_{BH}).
\eeq
Using the energy $E$ and the Bekenstein-Hawking energy to form the universe-size AdS hole
as given by $E_{BH}=R/(2G)$ to replace $S_B$ and $ S_{BH}$, (16) becomes
\beq 
S_H=\dpf{2\pi}{3\sqrt{3}}\sqrt{4E_{BH}R(3ER-E_{BH}R)}.
\eeq
If one takes the Virasoro operator as being $L_0=3ER$ and the central charge
as $c/6=4E_{BH}R$, we get 
\beq  
S_H=\dpf{2\pi}{3\sqrt{3}}\sqrt{c(L_0-c/24)/6},
\eeq
which is in striking resemblance with the Cardy formula in CFT.

The result of relating the Friedmann equation to the Cardy formula cannot be naively
extended to
AdS closed and open universes. If one supposes that there forms a universe-size AdS hole
satisfying eqs(11,12), one can learn directly from (10) that $H=0$ and $HR=\sqrt{2}$ for
$k=1$ and $k=-1$ AdS universes respectively. This corresponds to say that if closed and
open AdS universes are occupied by universe-size AdS black holes, self-gravity of the
system will always be limited or very strong for closed or open AdS universe
respectively. It is impossible to unify the Bekenstein bound and Hubble bound in these
AdS universes. On the other hand, if one supposes that a turning point $HR=1$ exists for
closed and open AdS universes, (10) will tell us that for a closed AdS universe
$R=(6GE/\Lambda)^{1/3}-2/(6GE\Lambda^2)^{1/3}$, while for the open AdS case
$R=(6GE/\Lambda)^{1/3}$, which is smaller or bigger than the AdS black hole radius
expressed in (11), respectively. This means that at the turning point $HR=1$, neither the
closed AdS universe nor the open AdS universe can accommodate a universe-size AdS
black hole in it. The general philosophy of the holography which needs the largest black
hole to fit inside the space to describe holography cannot be satisfied at the turning
point $HR=1$ of closed and open AdS universes. The fact that it is not possible either to
obtain a turning point between weak and strong self gravity, or to form an AdS black
hole with the size of the universe implies that there is no connection between the
Friedmann equation and the Cardy formula in closed or open AdS cosmologies.

In summary, we have investigated the relationship between Friedmann equation and Cardy
formula in de Sitter and AdS universes. We found that at the turning point, when the
universe evolves from limited self-gravity to strongly self-gravity system,
relation between Friedmann equation and entropy bounds matches strikingly the Cardy
formula
in de Sitter closed, AdS flat universes.
Our result generalized that obtained in closed FLRW universes without cosmological
constant [15]. This correspondence further supports holography, since the
Friedmann equation describing the bulk can be related to Cardy formula on the boundary
directly. The Cardy formula obtained unifies the Bekenstein bound and Hubble bound in de
Sitter and AdS cosmological models. It is worth noting that we did not specify the
equation of state of matter in our discussions, thus the result derived here is
independent of the matter state. The verification of the validity of the generalized
Cardy formula from coupled CFT is needed to be carried out in the future.

For de Sitter open or flat universes and AdS closed or open universes, the correspondence
between
Friedmann equation and Cardy formula has not been found. This is not so surprising if we
notice that the largest universe-size black hole cannot be formed in those universes. The
general philosophy of the holographic principle cannot be satisfied in those context. The
unification of Bekenstein bound and Hubble bound in those cases does not look possible.

ACKNOWLEDGEMENT: This work was partically supported by  
Fundac$\tilde{a}$o de Amparo $\grave{a}$ Pesquisa do Estado de
S$\tilde{a}$o Paulo (FAPESP) and Conselho Nacional de Desenvolvimento
Cient$\acute{i}$fico e Tecnol$\acute{o}$gico (CNPQ).  B. Wang would also  
like to acknowledge the support given by Shanghai Science and Technology  
Commission as well as NNSF, China under contract No. 10005004. R. K. Su's work was
supported by NNSF of China.

\end{document}